\documentclass[pra, twosides,preprintnumbers, twocolumn]{revtex4}
\usepackage{amssymb}

\usepackage{graphicx, amsmath, amssymb}

\begin{document}

\preprint{SB/F/06-339}

\title{Total Quantum Zeno effect and Intelligent States for a two level system in a
squeezed bath}
\author{D. Mundarain$^{1}$, M. Orszag$^{2}$ and J. Stephany$^{1}$}
\address{ \
${}^{1}$ Departmento de F\'{\i}sica, Universidad Sim{\'o}n
Bol{\'\i}var,
Apartado Postal 89000, Caracas 1080A, Venezuela \\
${}^{2}$ Facultad de F\'{\i}sica, Pontificia Universidad
Cat\'{o}lica de Chile, Casilla 306, Santiago, Chile}

\begin{abstract}
In this work we show that by frequent measurements of adequately chosen
observables, a complete suppression of the decay in an exponentially
decaying two level system interacting with a squeezed bath is obtained. The
observables for which the effect is observed depend on the the squeezing
parameters of the bath. The initial states which display Total Zeno Effect
are intelligent states of two conjugate observables associated to the
electromagnetic fluctuations of the bath.
\end{abstract}

\maketitle

\section{Introduction}

Frequent measurements modify the dynamics of a quantum system. This result
of quantum measurement theory is known as the quantum Zeno effect(QZE)\cite
{1,2,3,4,5} and has attracted much attention since first discussed. The QZE
is related to the suppression of induced transitions in interacting systems
or the reduction of the decay rate in unstable systems. It has been also
pointed out that the opposite effect, i.e the acceleration of the decay
process, maybe caused by frequent measurement too and this effect is known
as Anti-Zeno effect(AZE). The experimental observation of QZE in the early
days was restricted to oscillating quantum systems \cite{6} but recently,
both QZE and AZE were successfully observed in irreversible decaying
processes.\cite{7,8,9}.

The quantum theory of measurement predicts a reduction in the decay rate of
an unstable system if the time between successive measurements is smaller
than the Zeno time which in general is smaller than the correlation time of
the bath. The effect is universal in the sense that it does not depend on
the measured observable whenever the time between measurements is very
small. This observation does not preclude the manifestation of the Zeno
effect for larger times ( for example in the exponentially decaying regime )
for well selected observables in a particular bath.

Recently it was shown in Ref.( \cite{mund} ) that reduction of the decay
rate occurs in an exponentially decaying two level system when interacting
with a squeezed bath. In this case variations on the squeezing phase of the
bath may lead to the appearance of either Zeno or anti-Zeno effect when
continuously monitoring the associated fictitious spin. In this work we show
that for the same system it is possible to select a couple observables whose
measurement for adequately prepared systems lead to the total suppression of
the transitions i.e Total Zeno Effect.

This paper is organized as follows: In section (\ref{sec2}) we discuss some
general facts and review some results obtained in reference \cite{mund}
which are needed for our discussion. In Section (\ref{sec3}) we define the
system we deal with and identify the observables and the corresponding
initial states which are shown to display Total Zeno Effect. In section (\ref
{sec4}) we show that the initial states which show Total Zeno Effect are
intelligent spin states, i.e states that saturate the Heisenberg Uncertainty
Relation between the (fictitious) spin operators. Finally, we discuss the
results in Section (\ref{sec5}).

\section{Total Zeno effect in unstable systems}

\label{sec2}

Consider a closed system with Hamiltonian $H$ and an observable $A$ with
discrete spectrum. If the system is initialized in an eigenstate $%
|a_n\rangle $ of $A$ with eigenvalue $a_n$, the probability of survival in a
sequence of $S$ measurements, that is the probability that in all
measurements one gets the same result $a_n$, is
\begin{equation}
P_n(\Delta t,S) = \left( 1 - \frac{\Delta t^2}{\hbar^2} \Delta_n^2 \mathit{H}%
\right)^S
\end{equation}
where
\begin{equation}
\Delta_n^2 \mathit{H} = \langle a_n |\mathit{H}^2|a_n\rangle -\langle a_n |%
\mathit{H}|a_n\rangle^2
\end{equation}
and $\Delta t$ is the time between consecutive measurements. In the limit of
continuous monitoring ( $S \rightarrow \infty ,\Delta t \rightarrow 0$ and $%
S \Delta t \rightarrow t $ ), $P_n \rightarrow 1$ and the system is freezed
in the initial state.

For an unstable system and considering evolution times larger than the
correlation time of the reservoir with which it interacts, the irreversible
evolution is well described in terms of the Liouville operator $\mathit{L}%
\{\rho \}$ by the master equation
\begin{equation}
\frac{\partial \rho }{\partial t}=\mathit{L}\{\rho \}\ .  \label{master}
\end{equation}
If measurements are done frequently, the master equation (\ref{master}) is
modified. The survival probability is time dependent and is shown to be
given by \cite{mund}
\begin{equation}
P_{n}(t)=\exp \left\{ \langle a_{n}|\mathit{L}\{|a_{n}\rangle \langle
a_{n}|\}|a_{n}\rangle t\right\} \ \ .  \label{ec13}
\end{equation}
This expression is valid only when the time between consecutive measurements
is small enough but greater than the correlation time of the bath. For
mathematical simplicity in what follows we consider the zero correlation
time limit for the bath and then one is allowed to take the limit of
continuous monitoring. From equation (\ref{ec13}) one observe that the Total
Zeno Effect is possible when
\begin{equation}
\langle a_{n}|\mathit{L}\{|a_{n}\rangle \langle a_{n}|\}|a_{n}\rangle =0\ \ .
\label{ec1239}
\end{equation}
Then, for times larger than the correlation time, the possibility of having
Total Zeno effect depends on the dynamics of the system ( determined by the
interaction with the baths), on the particular observable to be measured and
on the initial state.

The concept of zero correlation time of the bath $\tau _{D}$ is of course an
idealization. If this time is not zero, the equation (\ref{ec13}) is only
approximate, since $\Delta t$ cannot be strictly zero and at the same time
be larger than $\tau _{D}$. Also, if equation (\ref{ec1239}) is satisfied,
then equation (\ref{ec13}) must be corrected, taking the next non-zero
contribution in the expansion of $\rho (\Delta t).$

In that case, a decay rate proportional to $\Delta t$ appears, and the decay
time is $\varpropto \frac{1}{\gamma ^{2}\Delta t}$, which is in general a
number much larger than the typical evolution time of the system. Notice
that as the spectrum of the squeezed bath gets broader, $\tau _{D}$ becomes
smaller, and one is able to choose a smaller $\Delta t$, approaching in this
way, the ideal situation and the Total Zeno Effect.

\section{Total Zeno observables}

\label{sec3}

In the interaction picture the Liouville operator for a two level system in
a broadband squeezed vacuum has the following structure \cite{gar},

\begin{eqnarray}
L\{\rho \} &=&\frac{1}{2}\gamma \left( N+1\right) \left( 2\sigma _{-}{\rho }%
\sigma _{+}-\sigma _{+}\sigma _{-}{\rho }-{\rho }\sigma _{+}\sigma
_{-}\right)  \nonumber  \label{em1} \\
&&\frac{1}{2}\gamma N\left( 2\sigma _{+}{\rho }\sigma _{-}-\sigma _{-}\sigma
_{+}{\rho }-{\rho }\sigma _{-}\sigma _{+}\right)  \nonumber \\
&&-\gamma Me^{i\phi }\sigma _{+}{\rho }\sigma _{+}-\gamma Me^{-i\phi }\sigma
_{-}{\rho }\sigma _{-}\ \ .
\end{eqnarray}
whereh $\gamma $ is the vacuum decay constant and $N,M=\sqrt{N(N+1)}$ and $%
\psi $ are the parameters of the squeezed bath. Here $\sigma _{-}$ and $%
\sigma _{+}$ are the two ladder operators,

\begin{equation}
\sigma_{+} =\frac{1}{2}(\sigma_x+i\sigma_y) \qquad \sigma_{-} =\frac{1}{2}%
(\sigma_x+i\sigma_y)
\end{equation}
with $\sigma _{x}$,$\sigma _{y}$ and $\sigma _{z}$ the Pauli matrices.

Let us introduce the Bloch representation of the two level density matrix
\begin{equation}
\rho =\frac{1}{2}\left( 1+\vec{\rho} \cdot \vec{\sigma}\right)  \label{dm1}
\end{equation}
In this representation the master equation takes the form:

\begin{eqnarray}
\frac{\partial \rho }{\partial t} &=&-\frac{1}{2}\gamma \left( N+1\right)
\left( (1+\rho _{z})\sigma _{z}+\frac{1}{2}\rho _{x}\sigma _{x}+\frac{1}{2}{%
\rho _{y}}\sigma _{y}\right)  \nonumber \\
&&+\frac{1}{2}\gamma N\left( (1-\rho _{z})\sigma _{z}-\frac{1}{2}\rho
_{x}\sigma _{x}-\frac{1}{2}{\rho _{y}}\sigma _{y}\right)  \nonumber \\
&&-\frac{1}{2}\gamma M\rho _{x}(\cos (\psi )\sigma _{x}-\sin (\psi )\sigma
_{y})  \nonumber \\
&&+\frac{1}{2}\gamma M\rho _{y}(\sin (\psi )\sigma _{x}+\cos (\psi )\sigma
_{y})
\end{eqnarray}
and has the following solutions for Bloch vector components that give the
behavior of the system without measurements
\begin{eqnarray}
\rho _{x}(t) &=&\left( \rho _{x}(0)\sin ^{2}(\psi /2)+\rho _{y}(0)\sin (\psi
/2)\cos (\psi /2)\right)  \nonumber \\
&&e^{-\gamma (N+1/2-M)\,t}  \nonumber \\
&&+\left( \rho _{x}(0)\cos ^{2}(\psi /2)-\rho _{y}(0)\sin (\psi /2)\cos
(\psi /2)\right)  \nonumber \\
&&e^{-\gamma (N+1/2+M)\,t}  \label{ec1222}
\end{eqnarray}
\begin{eqnarray}
\rho _{y}(t) &=&\left( \rho _{y}(0)\cos ^{2}(\psi /2)+\rho _{x}(0)\sin (\psi
/2)\cos (\psi /2)\right)  \nonumber \\
&&e^{-\gamma (N+1/2-M)\,t}  \nonumber \\
&&+\left( \rho _{y}(0)\sin ^{2}(\psi /2)-\rho _{x}(0)\sin (\phi /2)\cos
(\psi /2)\right)  \nonumber \\
&&e^{-\gamma (N+1/2+M)\,t}  \label{ec1223}
\end{eqnarray}
\begin{equation}
\rho _{z}(t)=\rho _{z}(0)e^{-\gamma (2N+1)t}+\frac{1}{2N+1}\left( e^{-\gamma
(2N+1)t}-1\right)
\end{equation}
Now consider the hermitian operator $\sigma _{\mu }$ associated to the
fictitious spin component in the direction of the unitary vector $\hat{\mu}%
=(\cos (\phi )\sin (\theta ),\sin (\phi )\sin (\theta ),\cos (\theta ))$
defined by the angles $\theta $ and $\phi $,
\begin{equation}
\sigma _{\mu }=\vec{\sigma}\cdot \hat{\mu}=\sigma _{x}\cos (\phi )\sin
(\theta )+\sigma _{y}\sin (\phi )\sin (\theta )+\sigma _{z}\cos (\theta )
\end{equation}
The eigenstates of $\sigma _{\mu }$ are,
\begin{equation}
|+\rangle _{\mu }=\cos (\theta /2)\,|+\rangle +\sin (\theta /2)\,\exp {%
(i\phi )}\,|-\rangle
\end{equation}
\begin{equation}
|-\rangle _{\mu }=-\sin (\theta /2)\,|+\rangle +\cos (\theta /2)\,\exp {%
(i\phi )}\,|-\rangle
\end{equation}

If the system is initialized in the state $|+\rangle _{\mu }$ the survival
probability at time $t$ is
\begin{equation}
P_{\mu }^{+}(t)=\exp \left\{ F(\theta ,\phi )\,\,t\,\right\}
\end{equation}
where
\begin{equation}
F(\theta ,\phi )=\,_{\mu }\langle +|\,\,L\left\{ \,\,|+\rangle _{\mu
}\,_{\mu }\langle +|\,\,\right\} \,\,|+\rangle _{\mu }\ \ .
\end{equation}
In this case the function $F(\theta ,\phi )$ has the structure
\begin{eqnarray}
F(\theta ,\phi ) &=&-\frac{1}{2}\gamma \left( N+1\right) \left( \rho
_{z}(0)+\rho _{z}^{2}(0)+\frac{1}{2}\rho _{x}^{2}(0)+\frac{1}{2}{\rho
_{y}^{2}(0)}\right)  \nonumber \\
&&+\frac{1}{2}\gamma N\left( (\rho _{z}(0)-\rho _{z}^{2}(0)-\frac{1}{2}\rho
_{x}^{2}(0)-\frac{1}{2}{\rho _{y}^{2}(0)}\right)  \nonumber \\
&&-\frac{1}{2}\gamma M\rho _{x}(0)(\cos (\psi )\rho _{x}(0)-\sin (\psi ){%
\rho _{y}(0)})  \nonumber \\
&&+\frac{1}{2}\gamma M\rho _{y}(0)(\sin (\psi )\rho _{x}(0)+\cos (\psi ){%
\rho _{y}(0)})
\end{eqnarray}
where now $\vec{\rho}(0)=\hat{\mu}$ is a function of the angles..

\begin{figure}[h]
\includegraphics[width=6cm,angle=-90]{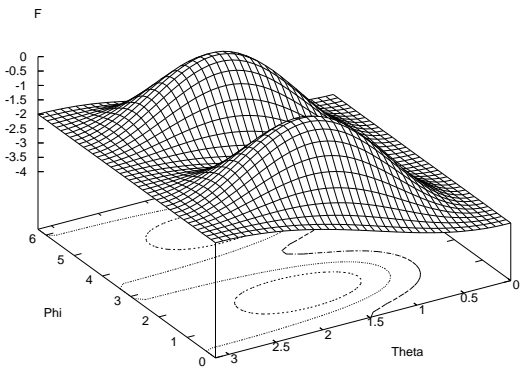}\bigskip
\caption{$F(\phi,\theta)$ for $N=1$ and $\psi =0$ }
\label{etiqueta1}
\end{figure}

In figure (\ref{etiqueta1}) we show $F(\phi ,\theta )$ for $N=1$ and $\psi
=0 $ as function of $\phi $ and $\theta $. The maxima correspond to $F(\phi
,\theta )=0$. For arbitrary values of $N$ and $\psi $ there are two maxima
corresponding to the following angles:

\begin{equation}
\phi_1^M = \frac{\pi-\psi}{2} \quad \mathrm{and} \quad \cos(\theta^M) = -
\frac{1}{2 \left( N+M+ 1/2\right)}
\end{equation}
and
\begin{equation}
\phi_2^M = \frac{\pi-\psi}{2}+\pi \quad \mathrm{and} \quad \cos(\theta^M) =
- \frac{1}{2 \left( N+M+ 1/2\right)}
\end{equation}
These preferential directions given by the vectors $\mathbf{\hat{\mu}_1}%
=(\cos(\phi_1^M)\sin(\theta^M),\sin(\phi_1^M)\sin (\theta^M),\cos(\theta^M))$
and $\mathbf{\hat{\mu}_2}=(\cos(\phi_2^M)\sin(\theta^M),\sin(\phi_2^M)\sin(%
\theta^M),\cos(\theta^M))$) define the operators $\sigma _{\mu_1}$ and $%
\sigma _{\mu_2}$ which show Total Zeno Effect if the initial state of the
system is the eigenstate $|+\rangle _{\mu_1}$ or respectively $|+\rangle
_{\mu_2}$.

To be specific let us consider measurements of the observable $%
\sigma_{\mu_1}= \vec{\sigma} \cdot \hat{\mu}_1$ (analogous results are
obtained for $\sigma_{\mu_2}$). The modified master equation with
measurements of $\sigma_{\mu_1}$ is given by \cite{mund},
\begin{equation}
\frac{\partial \rho }{\partial t}=P_{\mu _{1}}^{+}\,L\left\{ \rho \right\}
\,P_{\mu _{1}}^{+}+\left( 1-P_{\mu _{1}}^{+}\right) \,L\left\{ \rho \right\}
\,\left( 1-P_{\mu _{1}}^{+}\right)  \label{ecmcm}
\end{equation}
where
\begin{equation}
P_{\mu _{1}}^{+}=|+\rangle _{\mu _{1}}\,_{\mu _{1}}\langle +|
\end{equation}
and $\mathit{L}\{\rho \}$ is given by (\ref{em1}).

\begin{figure}[h]
\includegraphics[width=6cm,angle=-90]{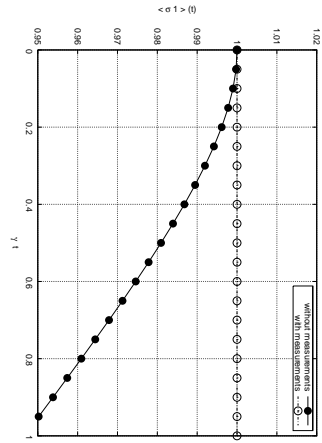}\bigskip
\caption{$\langle \sigma _{\mu _{1}}(t)\rangle $ for $N=1$ and $\psi =0$.
Solid circles: no measurements. Empty circles: with measurements. One
measures $\sigma _{\mu _{1}}$ and the initial state is $|+\rangle _{\mu
_{1}} $ }
\label{etiqueta2}
\end{figure}

Besides of the Total Zeno effect obtained in the cases specified above it is
also very interesting to discuss the effect of measurements for other
choices of the initial state. This can be done numerically. In figure (\ref
{etiqueta2}) we show the evolution of $\langle \sigma _{\mu _{1}}\rangle $,
that is the mean value of observable $\sigma _{\mu _{1}}$, when the system
is initialized in the state $|+\rangle _{\mu _{1}}$ without measurements
(master equation (\ref{em1})) and with frequent monitoring of $\sigma _{\mu
_{1}}$ (master equation (\ref{ecmcm})). Consistently with our discussion of
frequent measurements, the system is freezed in the state $|+\rangle _{\mu
_{1}}$ (Total Zeno Effect).

\begin{figure}[h]
\includegraphics[width=6cm,angle=-90]{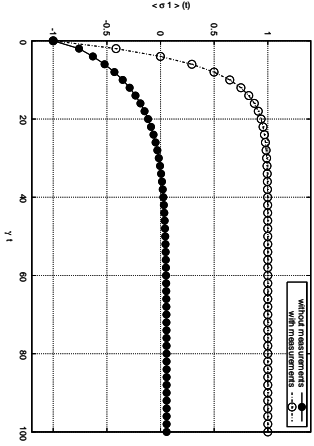}\bigskip
\caption{$\langle \sigma _{\mu _{1}}(t)\rangle $ for $N=1$ y $\psi =0$.
Solid circles: no measurements. Empty circles: with measurements. One
measures $\sigma _{\mu _{1}}$ and the initial state is $|-\rangle _{\mu
_{1}} $}
\label{etiqueta3}
\end{figure}

In figure (\ref{etiqueta3}) we show the time evolution of $\langle \sigma
_{\mu _{1}}\rangle $ when the initial state is $|-\rangle _{\mu _{1}}$
without measurements and with measurements of the same observable as in
previous case. One observes that with measurements the system evolves from $%
|-\rangle _{\mu _{1}}$ to $|+\rangle _{\mu _{1}}$. In general for any
initial state the system under frequent measurements evolves to $|+\rangle
_{\mu _{1}}$ which is the stationary state of Eq. ( \ref{ecmcm}) whenever we
do measurements in $\sigma _{\mu _{1}}$. Analogous effects are observed if
one measures $\sigma _{\mu _{2}}$. In contrast, for measurements in other
directions different from those defined by $\mathbf{\hat{\mu}_{1}}$ or $%
\mathbf{\hat{\mu}_{2}}$ , the system evolves to states which are not
eigenstates of the measured observables.

\section{Intelligent States}

\label{sec4}

Aragone et al \cite{ar} considered well defined angular momentum states that
satisfy the equality $(\Delta J_{x}\Delta J_{y})^{2}=\frac{1}{4}\mid \langle
J_{z}\rangle \mid ^{2}$ in the uncertainty relation. They are called
Intelligent States in the literature. The difference with the coherent or
squeezed states, associated to harmonic oscillators, is that these
Intelligent States are not Minimum Uncertainty States (MUS), since the
uncertainty is a function of the state itself.

In this section we show that the states $|+\rangle _{\mu _{1}}$ and $%
|+\rangle _{\mu _{2}}$ are intelligent states of two observables associated
to the bath fluctuations. The master equation (\ref{em1}) can be written in
an explicit Lindblad form
\begin{equation}
\frac{\partial \rho }{\partial t}=\frac{\gamma }{2}\left\{ 2S\rho S^{\dagger
}-\rho S^{\dagger }S-S^{\dagger }S\rho \right\}  \label{em2}
\end{equation}
using only one Lindblad operator $S$,
\begin{eqnarray}
S &=&\sqrt{N+1}\sigma _{-}-\sqrt{N}\exp \left\{ i\psi \right\} \sigma ^{+}
\nonumber \\
&=&\cosh (r)\sigma _{-}-\sinh (r)\exp \left\{ i\psi \right\} \sigma ^{+}
\end{eqnarray}
Obviously the eigenstates of $S$ satisfy the condition (\ref{ec1239}).
Moreover the states $|\phi _{1}\rangle =|+\rangle _{\mu _{1}}$ and $|\phi
_{2}\rangle =|+\rangle _{\mu _{2}}$ are the two eigenstates of $S$ with
eigenvalues $\lambda _{\pm }=\pm i\sqrt{M}\exp \{i\psi /2\}$.
\begin{equation}
S|\phi _{1,2}\rangle =\lambda _{\pm }|\phi _{1,2}\rangle
\end{equation}
Consider now the standard fictitious angular momentum operators for the two
level system are $\{J_{x}=\sigma _{x}/2,J_{y}=\sigma _{y}/2,J_{z}=\sigma
_{z}/2\}$ and also two rotated operators $J_{1}$ and $J_{2}$ which are
consistent with the electromagnetic bath fluctuations in phase space (see
fig. 2 in ref \cite{mund}) and which satisfy the same Heisenberg uncertainty
relation that $J_{x}$ and $J_{y}$ . They are,
\begin{eqnarray}
J_{1} &=&\exp \{i\psi /2J_{z}\}J_{x}\exp \{-i\psi /2J_{z}\}  \nonumber \\
&=&\cos (\psi /2)J_{x}-\sin (\psi /2)J_{y}
\end{eqnarray}
\begin{eqnarray}
J_{2} &=&\exp \{i\psi /2J_{z}\}J_{y}\exp \{-i\psi /2J_{z}\}  \nonumber \\
&=&\sin (\psi /2)J_{x}+\cos (\psi /2)J_{y}
\end{eqnarray}
These two operators are associated respectively with the major and minor
axes of the ellipse which represents the fluctuations of bath. Using Eqs. (%
\ref{ec1222}) and (\ref{ec1223}) one can show that the mean values of these
operators have the following exponentially decaying evolution:
\begin{equation}
\langle J_{1}\rangle (t)=\langle J_{1}\rangle (0)\exp \{-\gamma (N+M+1/2)t\}
\end{equation}

\begin{equation}
\langle J_{2}\rangle (t)=\langle J_{2}\rangle (0)\exp \{-\gamma (N-M+1/2)t\}
\end{equation}

Notice that the above averages decay with maximum and minimum rates
respectively.

In terms of $J_{1}$ y $J_{2}$ we have
\begin{equation}
J_{-}=\sigma =(J_{x}-iJ_{y})=\exp \{i\psi /2\}(J_{1}-iJ_{2})\ ,
\end{equation}
\begin{equation}
J_{+}=\sigma ^{\dagger }=(J_{x}+iJ_{y})=\exp \{-i\psi /2\}(J_{1}+iJ_{2})\ .
\end{equation}
$S$ has the form
\begin{equation}
S=\exp \{i\psi /2\}\left( \cosh (r)-\sinh (r)\right) \left( J_{1}-i\alpha
J_{2}\right)
\end{equation}
with
\begin{equation}
\alpha =\frac{\cosh (r)+\sinh (r)}{\cosh (r)-\sinh (r)}=\exp \{2r\}
\end{equation}

Following Rashid \textit{et al (}\cite{ra}\textit{)} we define a non
hermitian operator $J_{-}(\alpha )$
\begin{equation}
J_{-}(\alpha )=\frac{\left( J_{1}-i\alpha J_{2}\right) }{(1-\alpha
^{2})^{1/2}}
\end{equation}
so that
\begin{equation}
S=\exp \{i\psi /2\}\left( \cosh (r)-\sinh (r)\right) \,(1-\alpha
^{2})^{1/2}\,J_{-}(\alpha )
\end{equation}
After some algebra one obtains then that
\begin{equation}
S=2\,\lambda _{+}\,J_{-}(\alpha )
\end{equation}

The eigenstates of $S$ are then eigenstates of $J_{-}(\alpha )$ with
eigenvalues $\pm 1/2$. The eigenstates of $J_{-}(\alpha )$ are are also
shown to be intelligent states \textit{i.e} they satisfy the equality
condition in the Heisenberg uncertainty relation for $J_{1}$ and $J_{2}$ .
\begin{equation}
\Delta ^{2}J_{1}\Delta ^{2}J_{2}=\frac{|\langle J_{z}\rangle |^{2}}{4}
\end{equation}

The operator $J_{-}(\alpha )$ is obtained from $J_{1}$ by the following
transformation
\begin{equation}
J_{-}(\alpha )=\exp \{\beta J_{z}\}J_{1}\exp \{-\beta J_{z}\}
\end{equation}
with
\begin{equation}
\exp \{\beta \}=\sqrt{\frac{1-\alpha }{1+\alpha }}=i\sqrt{\frac{\sinh (r)}{%
\cosh (r)}}
\end{equation}
In terms of the real and imaginary parts of $\beta =\beta _{r}+i\beta _{i}$,
\begin{equation}
\beta _{i}=\frac{\pi }{2}
\end{equation}
\begin{equation}
\exp \{\beta _{r}\}=\sqrt{\frac{\sinh (r)}{\cosh (r)}}=\left( \frac{N}{N+1}%
\right) ^{1/4}
\end{equation}
$S$ takes the form,
\begin{eqnarray}
S &=&2i\sqrt{M}\exp \{i\psi /2\} \\
&&\,\exp \{i\frac{\pi }{2}J_{z}\}\,\exp \{\beta _{r}J_{z}\}\,J_{1}\,\exp \{-i%
\frac{\pi }{2}J_{z}\}\,\exp \{-\beta _{r}J_{z}\}  \nonumber
\end{eqnarray}
Finally $S$ may be written in the form:
\begin{equation}
S=2i\sqrt{M}\exp \{i\psi /2\}U\,J_{z}\,U^{-1}
\end{equation}
with
\begin{equation}
U=\,\exp \{i\frac{\pi }{2}J_{z}\}\,\exp \{\beta _{r}J_{z}\}\,\exp \{i\frac{%
\psi }{2}J_{z}\}\,\exp \{-i\frac{\pi }{2}J_{y}\}
\end{equation}
Then the eigenstates of $S$ could be obtained from the eigenstates of $J_{z}
$ using $U$ as
\begin{equation}
|\phi _{1,2}\rangle =C_{0}\text{ }U\text{ }|\pm \rangle
\end{equation}
where $C_{0}$ is a normalization constant. It is quite clear that $|\phi
_{1,2}\rangle $ are intelligent states of the observables $J_{1},J_{2},$%
which are rotated versions of $J_{x},J_{y}$. They are also quasi-intelligent
states of the original observables $J_{x},J_{y}$\cite{ra}).

One can verify the above result, by finding directly the eigenstates of $S$:
\begin{equation}
|\phi _{1,2}\rangle =\sqrt{\frac{N}{N+M}}|+\rangle \pm i\sqrt{\frac{M}{N+M}}%
e^{-i\psi /2}|-\rangle
\end{equation}

Finally, when the system is initialized in one of these states, the mean
value of $J_{1}$ is zero, $\langle J_{1}\rangle (t) =0$. Then, the term with
the biggest decaying rate does not appear in the mean value of the measured
observable $\sigma_{\mu}$ which becomes:
\begin{equation}
\langle \sigma_{\mu} \rangle (t) = \langle J_{2}\rangle (t) \sin (\theta^M)+
\langle J_{z}\rangle (t) \cos (\theta^M)
\end{equation}
Using the definition of the angle $\theta^M$ one can prove that
\begin{equation}
\frac{d \langle \sigma_{\mu} \rangle }{dt}(0) =0
\end{equation}
which is a neccesary condition in order to obtain Total Zeno Effect when one
is measuring the observable $\sigma_{\mu}$(See Fig (\ref{etiqueta2})).

\section{Discussion}

\label{sec5}

We have shown that Total Zeno Effect is obtained for two particular
observables $\sigma _{\mu_1}$ or $\sigma _{\mu_2}$, for which the azimuthal
phases in the fictitious spin representation depend on the phase of the
squeezing parameter of the bath and the polar phases depend on the squeeze
amplitude $r$. In this sense, the parameters of the squeezed bath specify
some definite atomic directions.

When performing frequent measurements on $\sigma _{\mu _{1}}$, starting from
the initial state $|+\rangle _{\mu _{1}}$, the system freezes at the initial
state as opposed to the usual decay when no measurements are done. On the
other hand, if the system is initially prepared in the state $|-\rangle
_{\mu _{1}},$ the frequent measurements on $\sigma _{\mu _{1}}$ will makes
it evolve from the state $|-\rangle _{\mu _{1}}$ to $|+\rangle _{\mu _{1}}$.
More generally, when performing the measurements on $\sigma _{\mu _{1}}$,
any initial state evolves to the same state $|+\rangle _{\mu _{1}}$ which is
the steady state of the master equation (\ref{ecmcm}) in this situation.

The above discussion could appear at a first sight surprising. However,
taking a more familiar case of a two-level atom in contact with a thermal
bath at zero temperature, if one starts from any initial state, the atom
will necessarily decay to the ground state. This is because the time
evolution of $\langle \sigma _{z}\rangle $ is the same with or without
measurements of $\sigma _{z}$. In both cases the system goes to the ground
state, which is an eigenstate of the measured observable $\sigma _{z}$. In
the limit $N,M\rightarrow 0$, $\sigma _{\mu _{1}}\rightarrow -\sigma _{z}$,
and the state $|+\rangle _{\mu _{1}}\rightarrow $ $|-\rangle _{z}$, which
agrees with the known results.

Finally, we also found that the eigenstates of $S$ are also
quasi-intelligent states of the observables $J_{x},J_{y}$ \textit{i.e}
intelligent states of the rotated version of the observables, that is, of $%
J_{1},J_{2}$. Starting from an eigenstate of $\sigma _{z}$, these
intelligent states are obtained by applying the transformation defined by $U$%
.

\subsection{Acknowledgements}

Two of the authors(D.M. and J.S.) were supported by Did-Usb Grant Gid-30 and
by Fonacit Grant No G-2001000712.

M.O was supported by Fondecyt \# 1051062 and Nucleo Milenio ICM(P02-049)

\bigskip

\end{document}